\documentclass[submission,copyright,creativecommons]{eptcs}
\usepackage[T1]{fontenc}
\usepackage[usenames,svgnames]{xcolor}
\usepackage{listings}
\usepackage[numbers,sort&compress]{natbib}
\usepackage{graphicx}
\usepackage{subfig}
\usepackage{url}
\usepackage{breakurl}
\usepackage{multirow,booktabs,array,longtable}
\usepackage[cmex10]{amsmath}
\usepackage{amssymb} 
\usepackage{hyperref}

\usepackage{nusmv}
\usepackage{xtend}
\providecommand{\event}{FESCA @ ETAPS 2014}

\def\fulltitle{Automated Mapping of UML Activity Diagrams to Formal Specifications for Supporting Containment Checking}
\def\shorttitle{Automated Mapping of UML Activity Diagrams to Formal Specifications}

\title{\fulltitle}
\author{Faiz UL Muram \qquad Huy Tran \qquad Uwe Zdun
\institute{Research Group Software Architecture\\University of Vienna, Austria.}
\email{faiz.ulmuram$|$huy.tran$|$uwe.zdun@univie.ac.at}}

\def\titlerunning{\shorttitle}
\def\authorrunning{F. U. Muram, H. Tran, and U. Zdun}

\usepackage[final,kerning=true,spacing=true,protrusion=true,expansion=true]{microtype}
\begin{document}
\maketitle

\begin{abstract}
Business analysts and domain experts are often sketching the behaviors of a software system using high-level models that are technology- and platform-independent. The developers will refine and enrich these high-level models with technical details. As a consequence, the refined models can deviate from the original models over time, especially when the two kinds of models evolve independently. In this context, we focus on behavior models; that is, we aim to ensure that the refined, low-level behavior models conform to the corresponding high-level behavior models. Based on existing formal verification techniques, we propose containment checking as a means to assess whether the system's behaviors described by the low-level models satisfy what has been specified in the high-level counterparts. One of the major obstacles is how to lessen the burden of creating formal specifications of the behavior models as well as consistency constraints, which is a tedious and error-prone task when done manually. Our approach presented in this paper aims at alleviating the aforementioned challenges by considering the behavior models as verification inputs and devising automated mappings of behavior models onto formal properties and descriptions that can be directly used by model checkers. We discuss various challenges in our approach and show the applicability of our approach in illustrative scenarios.
\end{abstract}

\section{Introduction}

Behavior models are used in many areas of software engineering. Examples of behavior models are UML activity diagrams, sequence diagrams, and state charts~\cite{UML} as well as BPMN business processes\cite{BPMN}, BPEL business processes \cite{BPEL}, Event-driven Process Chains (EPCs) \cite{Scheer:2000}, to name but a few. Many models are created as ``high-level'' models. That is, they are mainly used to convey the core concepts or principles of the reality they represent in an abstract and/or concise way. They are used for tasks such as defining core concepts and principles of a domain, enabling stakeholders to discuss them for instance in the course of a system design, or creating a common terminology. Also, high-level models are used as abstractions. For example, in design and architecture patterns high-level models are used as abstract representations of a best practice which need to be concretized in real systems. On the other hand, technical or ``low-level'' models are often created with purposes such as providing a precise specification of the source code, executing the model (e.g., in a process engine, interpreter, or virtual machine), or generating executable code directly from the model, e.g., in model-driven development (MDD). 

Unfortunately, the high-level models and their low-level counterpart are often drifting apart over time if, for example, contradictory specifications are defined. Therefore, it is important to ensure that the behaviors represented by the low-level models conform to what has been specified in the high-level models. In the literature, several techniques have been exploited to check the consistency and containment of different types of behavioral models~\cite{LucasM+2009}. Essentially, these approaches represent behavioral models in terms of formal specifications and/or properties and perform model checking to verify whether the formal specifications and properties are consistent. Most of these approaches assume that the formal specifications and properties are readily available or can be easily created beforehand. In reality, however, achieving such formal specifications is a tedious and challenging task because it often requires considerable knowledge on the leveraged formal techniques and it is often accomplished in a laborious, manual manner. 

In this paper, we propose a novel approach to alleviating the aforementioned problems. First, unlike most of existing approaches, we consider the high-level and low-level behavior models as inputs for checking containment. Thus, extra effort in creating consistency constraints that are widely used in the existing approaches can be reduced. Second, we devise an automated mapping of behavior models to formal specifications that can be used directly by model checkers for containment verification purpose. In particular, we derive primitive patterns for representing fundamental behavioral constructs such as sequences, parallel structures, and branches. These patterns are associated with formal description structures, for instance, linear temporal logic (LTL)~\cite{Pnueli:1977, Manna:1991} and the symbolic model language SMV~\cite{ClarkeM+1996,CimattiC+1999}. Based on model-driven transformation techniques, our approach enables automated generation of SMV descriptions and LTL formulas from given input behavioral models such as (in our case) UML activity diagrams~\cite{UML}. Using the NuSMV model checker\footnote{\url{http://nusmv.fbk.eu}}, we can assess whether a low-level technical model or implementation is consistent with the specification provided by a high-level, behavior model. In our work, we target UML activity diagrams because they are widely used in both industry and academia for representing software systems behavior. Linear temporal logic and SMV are used as the underlying formalism because, on the one hand, their expressiveness is suitable for model consistency and containment checking. On the other hand, we can leverage existing powerful tools for LTL and SMV model checking. Please note that our approach is also applicable for other behavioral models, such as the ones mentioned above, and can also be realized with different formal techniques with reasonable extra effort.  

The paper is structured as follows. In Section~\ref{sec:related-work}, we review the related approaches on supporting behavioral consistency checking in general and containment checking in particular. Our approach for automatically translating software behavior models onto formal properties and specifications is presented in detail in Section~\ref{sec:approach} along with illustrative examples. Afterwards, we discuss various challenges in supporting the aforementioned automated mappings in Section~\ref{sec:discussion}. Finally, we summarize the main contributions and discuss the planned future work in Section~\ref{sec:conclusion}.

\section{Related Work} \label{sec:related-work}

A considerable amount of consistency checking approaches have been proposed so far. Finkelstein et al. present an approach using temporal logic for checking the consistency of different viewpoints~\cite{FinkelsteinG+1994}. Similar approaches targeting different kinds of models or model checking techniques exist~\cite{LucasM+2009}. While some of those focus on behavior models, the major difference to our work is that those other approaches mainly consider the consistency between different kinds of models (or models and other representations of the same reality such as the implementations or requirements) but not the consistency of the same model at different abstraction levels. That is, we focus on checking the consistency of the containment of the high-level model in the low-level model, rather than checking the consistency of elements of two different representations. According to the methodology to handle consistency problems proposed by Engels et al.~\cite{EngelsK+2001}, we address the ``vertical consistency''---in contrast to ``horizontal consistency'' which checks consistency between models. 

Additionally, most of the existing approaches require the specification of a consistency condition for each consistency problem addressed. The goal of our approach in contrast is to use a high-level model, which is automatically translated to formal descriptions, as the specification of the consistency problem and check the low-level model whether it contains the high-level specifications.

There are only a few approaches for containment checking that use high-level models as inputs but they merely focus on structural models. Egyed introduces an approach based on structural transformation rules and transitive reasoning to check whether an UML class diagram conforms to another more abstract class diagram~\cite{Egyed2002}. Unfortunately, this approach alone does not suffice for the behavior models addressed in our work because we must assume that behavior models contain control structures, such as sequences, links, parallel gateways, and so on, which cannot be matched by structure only. 

Many other consistency checking approaches focus on different types of consistency but containment. Van Der Straeten et al. check the consistency of different UML models using a description logic based approach~\cite{StraetenM+2003}. This approach does not focus on consistency of high-level and low-level models, but rather on model-model, model-instance, and instance-instance conflicts. Ehrig and Tsiolakis use attributed graphs and graphical constraints to check the consistency of UML class and sequence diagrams~\cite{TsiolakisEhrig2000}. In particular, the existence, correct multiplicity, and valid scoping of model elements are checked. Graaf and van Deursen introduce a model-driven consistency checking approach for checking the consistency of various behavior models among each other~\cite{GraafVanDeursen2007}. The approach first normalizes the input models, then performs an automated model transformation to a state machine, and then compares the different state machines to detect inconsistencies.

There are a number of efforts on translating manually described constraints to formal representations. For instance, Czarnecki and Pietroszek check the well-formedness of feature diagrams using OCL constraints (i.e. structural models)~\cite{CzarneckiPietroszek2006}. Engels et al. check contracts between Web services and business processes (i.e., consistency between behavior and structural models)~\cite{EngelsG+2008}. Campbell et al. check for and visualize errors in UML diagrams~\cite{CampbellC+2002}. K{\"o}hler et al. introduce an approach to verifying a business process model and its implementation but merely concentrate on checking the termination property~\cite{KoehlerT+2002}. Eshuis considered using NuSMV for checking data integrity constraints between an activity diagram and a set of class diagrams~\cite{Eshuis2006}.

We note that, in most of the aforementioned approaches, the availability of the input formal specifications is often taken for granted. Unfortunately, achieving the formal specifications as inputs for model checking techniques is challenging and error-prone for the majority of developers as reasonable knowledge of formal languages and verification techniques is always necessary. Our approach aims at filling this gap by introducing primitive behavior patterns grounded on formal expressions that can support the automated mapping of the high-level and low-level behavior models to formal specifications.

\section{Approach} \label{sec:approach}

In this section, we firstly present an overview of our approach focusing on the automated mapping of behavior models into formal descriptions for containment checking. Afterwards, we elaborate on our proposed techniques for mapping high-level behavior models onto formal property specifications (i.e. LTL formulas in our case) and the low-level counterparts into formal SMV specifications. 

\subsection{Approach overview}\label{sub-sec:approach_overview}

Our containment checking approach addresses the consistency problems existing between high-level and low-level behavior models. The low-level counterparts are often resulting from various steps of refining and enriching the high-level models. In most of the cases, models evolve, often independently, over time. For instance, high-level models are changed according to new requirements, and low-level models are changed as the implementation is modified. As a consequence, the evolved models may include inconsistencies, such as the order of activities swaps during refinement of one of the models, some new activities are added in the high-level model but are not present in the low-level model, or some low-level activities are deleted without updating the high-level model.

The main goal of checking containment is to verify whether the behavior described by a low-level model still satisfies the behavior specified by a corresponding high-level model. In other words, {\em containment checking will assess if the execution traces produced by the low-level model  contain those produced by the high-level counterpart}. Assuming that the low-level behavior model is represented by a formal specification, we could achieve containment checking by checking that the high-level model's desired properties are satisfied by this formal specification. As a result, a model checker can be used to answer whether the formal specification satisfies these properties, i.e., the low-level behavior model conforms to the corresponding high-level model. Unfortunately, achieving formal representations and properties of behavior models under consideration is a challenging and error-prone task because it requires deep knowledge of the formal verification techniques.

\begin{figure}[htp]  
\centering
\includegraphics[scale=0.8]{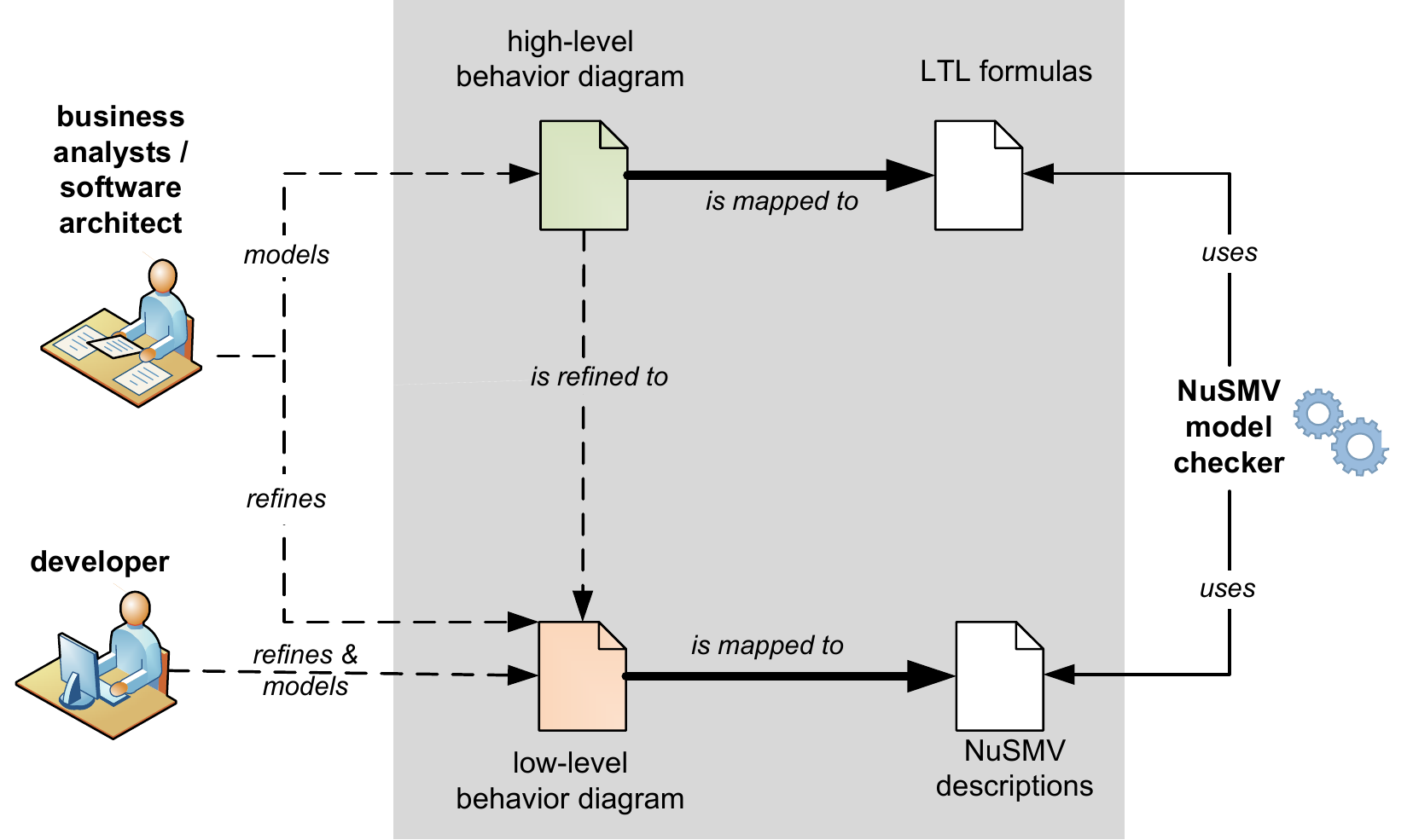}  
\caption{Overview of the approach} \label{fig:Approach_Overview} 
\end{figure}

Therefore, we present an approach to address this challenge by supporting automated translation of behavior models into formal specifications and properties that can be used by model checkers for containment checking. The main focus of our approach is represented by the grey box shown in Figure~\ref{fig:Approach_Overview}. In particular, we introduce primitives for representing fundamental behavioral constructs, such as sequences, parallel structures, and branches, in terms of LTL formulas. These primitives will be used to transform high-level UML activity diagrams into LTL properties. The low-level UML activity diagrams under consideration will be transformed to symbolic formal specifications. We note that these transformations are represented as solid arrows in Figure~\ref{fig:Approach_Overview} as they will be performed automatically. 

Finally, containment checking is performed on these models by using the NuSMV model checker, which supports symbolic model verification~\cite{ClarkeM+1996}. By analyzing the verification results reported by the NuSMV model checker, one can assess whether the input behavior models are consistent as well as reason about their inconsistency (if any exists) through the generated counterexamples. Verification and reasoning with the NuSMV model checker is, however, beyond the scope of this paper and will be part of our follow-on endeavor. 
  
The aforementioned concepts have been realized in our prototypical implementation. In our implementation, we exploit model-driven development techniques to support automated transformations of UML activity diagrams. In particular, we leverage Eclipse Xtend framework\footnote{\url{http://www.eclipse.org/xtend/documentation/2.4.3/Documentation.pdf}}, which provides  powerful and expressive languages and techniques for defining and executing transformation rules. As a result, our implementation can be easily integrated in the Eclipse development environment. 

\subsection{Transformation of High-level UML Activity Diagrams into LTL Formulas}\label{sub-sec:LTL_formalization} 

The first aspect of our approach is to map a set of fundamental behavioral constructs of UML activity diagrams, such as sequences, parallel structures, and branches,  into LTL-based properties. LTL is an expressive formalism that is widely used in formal verification tools. Formulas in LTL are usually constructed from boolean predicates and logical operators such as {\em not} (!), {\em or} (|), {\em and} (\&), {\em xor}, and {\em implication} ($\to$).  LTL also supports the specification of future with temporal operators such as `{\em always}' (G), `{\em next}' (X), `{\em eventually}' (F), and so on. For more details on LTL formulas as well as their semantics, please refer to~\cite{Pnueli:1977}. 
 
According to the OMG UML 2.4 specification~\cite{UML}, UML activity diagrams provide several constructs for describing the behavior of software systems. Without loss of generality, we consider a set of fundamental elements and control structures such as activities, actions, sequences, parallel structures (fork and join), branching (decision and merge) in the scope of this paper. Other constructs are also applicable using our approach in similar manner with little extra effort. The {\em xor} operator is often used in the LTL-based primitive for representing decision~\cite{BrambillaDSV05}. 

\begin{table}[ht]
{\small
\caption{Representation of UML activity diagram's elements using LTL primitives} \label{tab:Table_LTL_activity}
\begin{tabular}{|m{.32\textwidth}|m{.31\textwidth}|m{.28\textwidth}|}
\hline
{\bf Description} & {\bf Modeling Notation} & \bf LTL Primitive 
\\ 
\hline
{\bf Sequence}: A set of actions executed in sequential order. E.g., the action A1 will be performed before A2. 
& 
\begin{minipage}{.3\textwidth}
\includegraphics[width=4.2em]{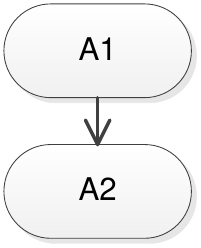}
\end{minipage}
& 
\begin{lstlisting}[language=SMV,xleftmargin=0em,aboveskip=-2pt,belowskip=-3ex,basicstyle=\ttfamily\small]
G (A1 -> F A2)
\end{lstlisting}
\\\hline
{\bf Fork Node}: The execution of a fork node leads to the parallel execution of subsequent actions (B1...Bn). 
&
\begin{minipage}{.3\textwidth}
\includegraphics[width=12em]{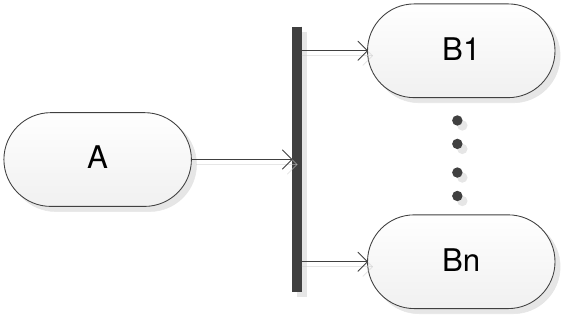}
\end{minipage}
& 
\begin{lstlisting}[language=SMV,xleftmargin=0em,aboveskip=-2pt,belowskip=-3ex,basicstyle=\ttfamily\small]
G (A -> (F B1 
        & F B2 
        & ... 
        & F Bn))
\end{lstlisting}
\\\hline
{\bf Join Node}: The execution of two or more parallel actions leads to the execution of Join Node.

&  
\begin{minipage}{.3\textwidth}
\includegraphics[width=12em]{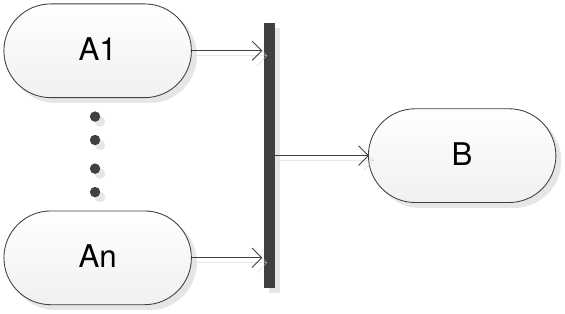}
\end{minipage}
& 
\begin{lstlisting}[language=SMV,xleftmargin=0em,aboveskip=-3pt,belowskip=-2ex,basicstyle=\ttfamily\small]
  (G (A1  
   &  A2  
   & ... 
   &  An) -> F B)
\end{lstlisting}
\\\hline
{\bf Decision Node}: The execution of a Decision Node eventually followed by the execution of one and only one action among the available set of actions based on the decision guard. 
& 
\begin{minipage}{.31\textwidth}
\includegraphics[width=13em]{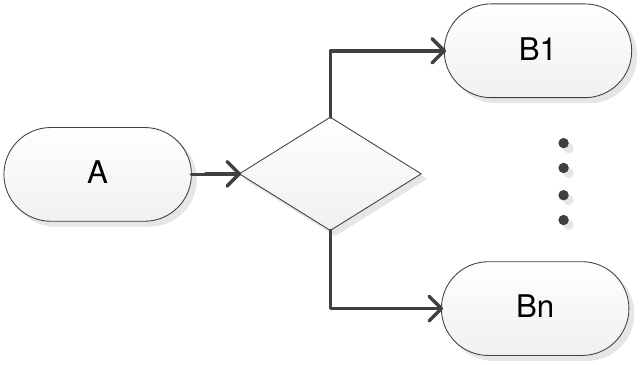}
\end{minipage}
& 
\begin{lstlisting}[language=SMV,xleftmargin=0em,aboveskip=-3pt,belowskip=-2ex,basicstyle=\ttfamily\small]
G (A -> (F B1 xor F B2 xor ...  xor F Bn))
\end{lstlisting}
\\\hline
{\bf Merge Node}: At least one action among a set of alternative execution of actions will lead to the execution of Merge Node.
& 
\begin{minipage}{.31\textwidth}
\includegraphics[width=13em]{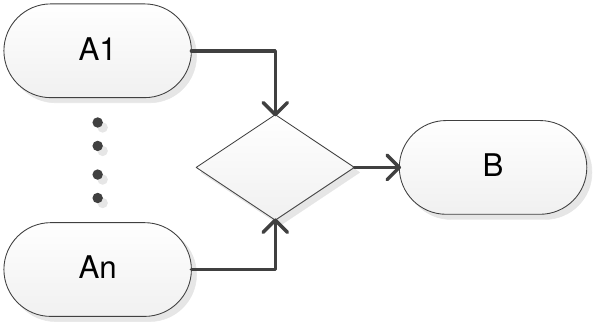}
\end{minipage}
&
\begin{lstlisting}[language=SMV,xleftmargin=0em,aboveskip=-3pt,belowskip=-2ex,basicstyle=\ttfamily\small]
G (A1 | A2 | ... | An -> F B)
\end{lstlisting}
\\\hline
\end{tabular}
} 
\end{table}

In Table~\ref{tab:Table_LTL_activity}, we present these constructs of UML activity diagrams and their informal description extracted from the UML specification~\cite{UML}. The right-hand side column contains the corresponding LTL-based primitive patterns for formally representing these constructs. Using these patterns, we traverse through the input UML activity diagram (i.e. the high-level model) and translate its elements into corresponding LTL formulas.

\begin{figure}[htp]  
\centering\addtolength{\abovecaptionskip}{-.6em}
\includegraphics[scale=0.85]{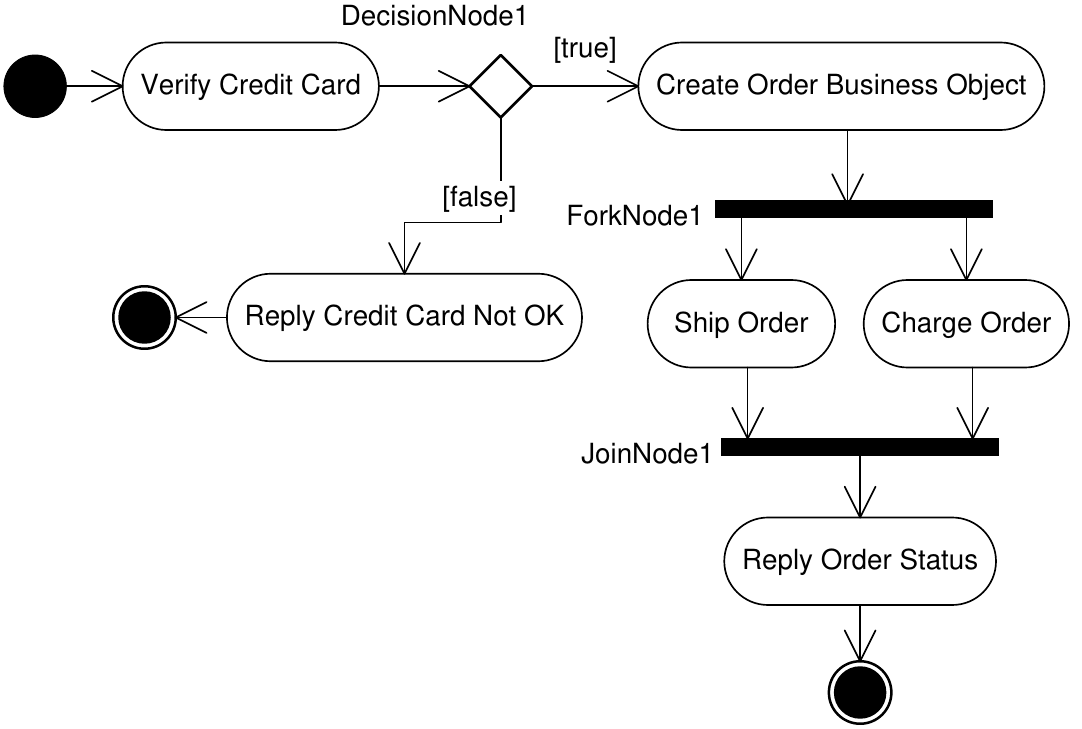}  
\caption{High-level UML activity diagram}\label{fig:HighLevel_Activitydiagram}
\end{figure}

We use a simplified order processing system example to demonstrate the transformation of UML activity diagrams to LTL formulas. The high-level model of the order processing system is shown in Figure~\ref{fig:HighLevel_Activitydiagram} and it includes sequential actions, a decision node, a fork node, and a join node.  Using the primitives shown in Table~\ref{tab:Table_LTL_activity}, we are able to generate the corresponding LTL formulas in Listing~\ref{GeneratedLTL} out of the aforementioned UML activity diagram.

\begin{lstlisting}[language=SMV,numbers=left,frame=tlrb,caption={The formal property specifications automatically generated from the high-level UML activity diagram},label=GeneratedLTL]
LTLSPEC G (InitialNode1 -> F VerifyCreditCard)
LTLSPEC G (VerifyCreditCard -> F ReplyCreditCardNotOK xor F CreateOrderBusinessObject)
LTLSPEC G (ReplyCreditCardNotOK -> F ActivityFinalNode1)
LTLSPEC G (CreateOrderBusinessObject -> F ShipOrder & F ChargeOrder)
LTLSPEC (G (ShipOrder & ChargeOrder) -> F ReplyOrderStatus)
LTLSPEC G (ReplyOrderStatus -> F ActivityFinalNode2)
\end{lstlisting} 

The automated transformation of high-level UML activity diagrams into LTL formulas has been realized using Eclipse Xtend. An excerpt of the Xtend template expression for describing the transformation of basic constructs such as sequence and fork node into corresponding LTL formulas is shown in Listing~\ref{uml_to_LTL}. Xtend templates expressions are multi-line strings, that are surrounded by triple single quotes ('''). These strings are further enhanced with expressions embraced in a pair of guillemets (i.e., ``\guillemotleft'' and ``\guillemotright''). Such expressions will be bound and evaluated according to the elements of the input UML activity diagram. 

The first template presents the mapping of actions into LTL formulas that are executed in the sequence. The second template defines generation of LTL formulas for the fork node. To this, first we get the incoming flow that triggers the fork node and map them into LTL rule. Afterward, the subsequent action(s) that are executed by the fork node are mapped into LTL rules. If there are more than one outgoing actions then `{\tt \& F}' is attached with every subsequent actions. These LTL formulas are generated according to the primitives presented in Table~\ref{tab:Table_LTL_activity}.

\begin{lstlisting}[language=Xtend,frame=tlrb,caption={Transformation of UML structures into LTL formulas},label=uml_to_LTL]
/* For the generation of sequential actions */
«IF (!convert(a).nullOrEmpty)»LTLSPEC G («a.name» -> F «convert(a)»);«ENDIF»
/* Template expression for the generation of a fork node */
«IF (!convert(a).nullOrEmpty)»LTLSPEC G («a.incomings.iterator().next().normalize» -> F «convert(a)»);
«ENDIF»
/* Create a Fork's guard condition */
 a.outgoings.forEach [ edge, i |
			val target = edge.target
			if (edge != null && target != null) {
						if (i > 0) 
							guard.append( " & F ")
						guard.append(target.name)
			}
	]
\end{lstlisting}

\subsection{Transformation of Low-level UML Activity Diagrams into SMV specifications}\label{sub-sec:low_level}

The second part of our approach automates the translation of low-level UML activity diagrams into formal SMV specifications. These specifications then can be used by the NuSMV model checker to verify against the LTL rules generated from the high-level counterparts to see whether the containment relationship between these diagrams is satisfied. The transformation templates for the  UML activity diagram constructs considered in this paper are shown in Table~\ref{tbl:uml-to-smv}. 

The transformation of a low-level UML activity diagram into SMV specification works as follows. Essentially, every node of a UML activity diagram except a decision node will be represented by a corresponding symbolic variable declared in the `{\tt VAR}' section. A decision node, as specified in UML 2.4 specification~\cite{UML}, will trigger the execution of one of its branching nodes according to their guarded conditions. Therefore, we map a decision node into a scalar variable. The value of the scalar variable is defined based on the values of the branch guarded conditions plus the constant `undetermined' used as initial state. 

After having nodes represented as symbolic variables, we need to map the control flow of the UML activity diagram into the corresponding state transition rules for each variable. The general description of a variable's state transitions is a combination of the statement {\tt init()}--defining the initial state--and {\tt next()}--defining the next state. As a result, given a certain node $n$ of a UML activity diagram represented by a NuSMV symbolic variable $a$, we need to define a set of statements as follows: ``{\tt init(a) := initial\_value; next(a) := next\_value;}''.

The {\tt next\_value} can be defined by a concrete value straightforwardly or through an exclusive choice of various possible values with respect to some constraints or previous states. In the later case, we can use the NuSMV construct ``{\tt case...esac}'' to specify such exclusive choices. Thus, the most plausible form of the state transitions is to (1) initialize the variable's state as {\tt FALSE}; (2) define a guard condition that trigger the execution of the node according to the UML specification~\cite{UML} such that the variable's state can become {\tt TRUE}; (3) define another choice to switch the variable's state back to {\tt FALSE} if it was {\tt TRUE} before that. These state transitions can be described in the following NuSMV code.

\begin{lstlisting}[language=SMV,backgroundcolor=\color{bgcolor}]
  init(a) := initial_value;       -- (1)
  next(a) := case                
		guard_condition : TRUE;  -- (2)
		a : FALSE;														-- (3)
  esac;
\end{lstlisting}

\begin{table}[htp]
\centering
\begin{tabular}{|m{.2\textwidth}|m{.74\textwidth}|}
\hline
{\bf UML constructs} & {\bf Transformation Rules}  \\ 
\hline
Initial Node $a$
& 
\begin{lstlisting}[language=SMV,	aboveskip=2pt,belowskip=-3ex,nolol]
  init(a) := TRUE;
  next(a) := case
    a : FALSE;
    TRUE    : a; 
  esac;
\end{lstlisting}
\\\hline
Final Node, Action, Fork Node, Join Node $a$ &  
\begin{lstlisting}[language=SMV,	aboveskip=2pt,belowskip=-3ex,nolol]
  init(a) := FALSE;
  next(a) := case  
    incoming_1 & incoming_2 & ... & incoming_n : TRUE;
    a : FALSE;
  esac;
\end{lstlisting}
\\\hline
Decision Node $a$ & 
\begin{lstlisting}[language=SMV,	aboveskip=2pt,belowskip=-3ex,nolol]
 init(a) := undetermined;
 next(a) := case
   incoming_1 & incoming_2 & ... & incoming_n$ : {outgoing_1, outgoing_2, ..., outgoing_n};
   a != undetermined : undetermined;
 esac;
\end{lstlisting}
 \\	\hline
Merge Node & 
\begin{lstlisting}[language=SMV,	aboveskip=2pt,belowskip=-3ex,nolol]
 init(a) := FALSE;
 next(a) := case
   incoming_1 | incoming_2 | ... | incoming_n : TRUE;
   a : FALSE;
 esac;						
\end{lstlisting}
\\\hline
\end{tabular}
\caption{Transformation of UML structures into SMV descriptions} \label{tbl:uml-to-smv}
\end{table}

The mapping rules shown in Table~\ref{tbl:uml-to-smv} define the state transition for different types of nodes of a UML activity diagram. The execution of the nodes of an activity diagram is based on token semantics~\cite{UML} similar to Petri Nets~\cite{Murata1989}. The {\tt Initial Nodes} are the starting points for a UML activity diagram, therefore, their initial state will be {\tt TRUE}. The initial state of other nodes (except Decision Node) are {\tt FALSE}. Apart from the {\tt Initial Nodes} that have no incoming edges, other nodes will be triggered with respect to their incoming control flows. We note that the UML specification denotes an ``implicit join'' in case a node has multiple incoming edges. Therefore, the guard for transitioning to the next state for a {\tt Final Node}, {\tt Action}, {\tt Decision Node}, or {\tt Fork Node} is similar to the default semantics of an explicit {\tt Join Node}, i.e., an ``{\em and}''-join of all tokens going through the incoming control flows. We use the operator logical {\em and} (``{\tt \&}'') of NuSMV to represent the {\em and}-join guard. Please note that the non-default ``{\tt joinSpec}'' of an explicit {\tt Join Node} can also be easily supported by altering the logical condition of the {\tt Join Node} in NuSMV. 

A {\tt Merge Node} is a special case, as it brings together multiple alternative flows. Therefore, we use the operator {\em or} ``{\tt |}'' of NuSMV to describe the guard condition of a {\tt Merge Node}. The possible states of a symbolic variable representing a node (except Decision Node) are {\tt TRUE} and {\tt FALSE}. In case of a {\tt Decision Node}, the initial state is predefined as {\tt undetermined} and the possible next states is an exclusive choice of its outgoing branches. We use the notion of {\em non-deterministic assignments} in NuSMV to describe the outcome of a {\tt Decision Node} such that the NuSMV model checker can explore all possible outcomes for verification. We have implemented these rules using the Eclispe Xtend language and exploit the code generation template of Xtend to automatically produce NuSMV descriptions out of a low-level UML activity diagram. 

\begin{figure}[htp]  
\centering
\includegraphics[scale=0.8]{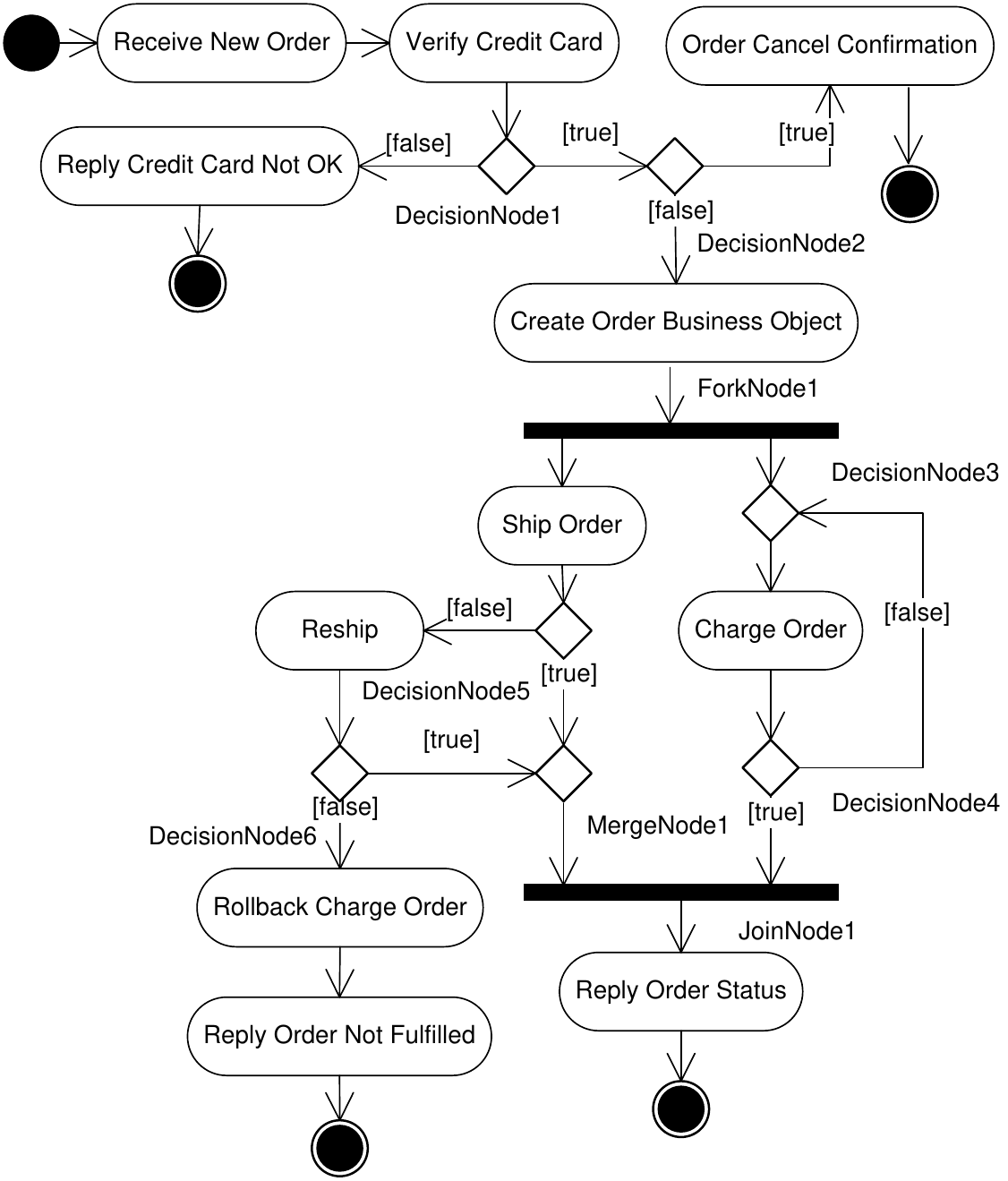}  
\caption{Low-level UML activity diagram for the order processing system\label{fig:Low_level}}  
\end{figure}

To illustrate the transformation of low-level UML activity diagram into NuSMV, we use one of the refined versions of the order processing system mentioned above. The low-level model of the order processing system is shown in Figure~\ref{fig:Low_level}. In this diagram, technical details are added that are required for an implementation of the order processing system, but have been omitted by the domain experts. 

\noindent\begin{lstlisting}[language=SMV, xleftmargin=\parindent,frame=tblr,backgroundcolor=\color{bgcolor},caption={NuSMV symbolic variables for representing UML activity diagram nodes},label=lst-variables]
MODULE
VAR
	InitialNode1 : boolean;
	ReceiveNewOrder : boolean;
	VerifyCreditCard : boolean;
	DecisionNode1 : {undetermined, guard_DecisionNode1_ReplyCreditCardNotOK, guard_DecisionNode1_DecisionNode2};
	DecisionNode2 : {undetermined, guard_DecisionNode2_ConfirmOrderCancelation, guard_DecisionNode2_CreateOrderBusinessObject};
  -- partially omitted
\end{lstlisting}

The transformation rules explained above can be used to automatically produce a corresponding NuSMV description. First, the nodes of the activity diagram (except decision nodes) are represented by a boolean symbolic variable as shown in Listing~\ref{lst-variables}. Because a decision node can assign the output token to one of the outgoing branches, we use a scalar variable to represent a decision node. The range of values of the scalar variable correspond to the guards of the outgoing branches of the decision node. Please note that we have given names for all structural nodes in Figure~\ref{fig:Low_level} to make the example mapping to the NuSMV description below easily understandable. This is not necessary in our approach, our code generator could also automatically assign names.

We show in Listing~\ref{lst-state-transitions} an excerpt of the corresponding state transitions of the aforementioned symbolic variables according to the rules in Table~\ref{tbl:uml-to-smv}. Some repetitive excerpts, for instance, produced from the similar types of nodes, have been omitted due to space limitation.

\begin{lstlisting}[language=SMV,frame=tblr,xleftmargin=10pt,backgroundcolor=\color{bgcolor},caption={NuSMV state transitions for the low-level UML activity diagram},label=lst-state-transitions]
ASSIGN
init(InitialNode1) := TRUE;
next(InitialNode1) := case
        InitialNode1 : FALSE;
    esac;						
init(ReceiveNewOrder) := FALSE;
next(ReceiveNewOrder) := case
        InitialNode1 : TRUE;
        ReceiveNewOrder : FALSE;
    esac;
...
init(DecisionNode1) := undetermined;
next(DecisionNode1) := case
        VerifyCreditCard : {guard_DecisionNode1_ReplyCreditCardNotOK, guard_DecisionNode1_DecisionNode2};
        DecisionNode1 != undetermined : undetermined;
    esac;
init(ReplyCreditCardNotOK) := FALSE;
next(ReplyCreditCardNotOK) := case
        (DecisionNode1 = guard_DecisionNode1_ReplyCreditCardNotOK) : TRUE;
        ReplyCreditCardNotOK : FALSE;
    esac;
init(DecisionNode2) := undetermined;
next(DecisionNode2) := case
        (DecisionNode1 = guard_DecisionNode1_DecisionNode2) : {guard_DecisionNode2_ConfirmOrderCancelation, guard_DecisionNode2_CreateOrderBusinessObject};
        DecisionNode2 != undetermined : undetermined;
    esac;
init(ActivityFinalNode1) := FALSE;
next(ActivityFinalNode1) := case
        ReplyCreditCardNotOK : TRUE;
        ActivityFinalNode1 : FALSE;
    esac;
init(ConfirmOrderCancelation) := FALSE;
next(ConfirmOrderCancelation) := case
        (DecisionNode2 = guard_DecisionNode2_ConfirmOrderCancelation) : TRUE;
        ConfirmOrderCancelation : FALSE;
    esac;
init(CreateOrderBusinessObject) := FALSE;
next(CreateOrderBusinessObject) := case
        (DecisionNode2 = guard_DecisionNode2_CreateOrderBusinessObject) : TRUE;
        CreateOrderBusinessObject : FALSE;
    esac;
-- generated code is partially omitted
init(ForkNode1) := FALSE;
next(ForkNode1) := case
        CreateOrderBusinessObject : TRUE;
        ForkNode1 : FALSE;
    esac;						
init(ShipOrder) := FALSE;
next(ShipOrder) := case
        ForkNode1 : TRUE;
        ShipOrder : FALSE;
    esac;
init(MergeNode2) := FALSE;
next(MergeNode2) := case
        (DecisionNode4 = guard_DecisionNode4_MergeNode2) | ForkNode1 : TRUE;
        MergeNode2 : FALSE;
    esac;						
init(DecisionNode5) := undetermined;
next(DecisionNode5) := case
        ShipOrder : {guard_DecisionNode5_MergeNode1, guard_DecisionNode5_Reship};
        DecisionNode5 != undetermined : undetermined;
    esac;
-- generated code is partially omitted
\end{lstlisting}

\subsection{Illustration of Containment Checking Using NuSMV}\label{sub-sec:containment}
 
In this section, we illustrate how the LTL properties and NuSMV descriptions generated in the previous steps can be used for containment checking. We use the NuSMV model checker to perform containment checking based on the formal SMV specifications illustrated in Figure~\ref{fig:Low_level} and the LTL rules generated from the high-level counterparts presented in Listing \ref{GeneratedLTL}. The containment checking result is shown in Listing~\ref{CounterExample}. By looking at the result, we see that the generated NuSMV description satisfies all LTL properties except ``{\tt LTLSPEC G (VerifyCreditCard -> F ReplyCreditCardNotOK xor F CreateOrderBusinessObject)}''. This unsatisfied condition is due to the occurrence of the {\tt Decision Node 2} that leads to two possibilities outcomes, which are {\tt CreateOrderBusinessObject} and {\tt OrderCancelConfirmation}. In other words, the verification result indicates that the descriptions of the low-level order processing model do not conform the LTL formulas generated from the high-level counterparts. According to the specification, ``ReplyCreditCardNotOK'' should becomes true in the future. But in the counter example, possible loop is presented and marked with the ``Loop starts here'' line. The line indicates that the ``ReplyCreditCardNotOK'' never becomes true in the loop, so the behavior can occur repeatedly. Hence, this LTL specification is false.

\begin{lstlisting}[basicstyle=\ttfamily\scriptsize,frame=tlrb, breaklines=true,xleftmargin=\parindent,backgroundcolor=\color{bgcolor},caption={NuSMV containment checking result along with a counterexample}, label=CounterExample]
$ NuSMV LowlevelOrderProcessingNotSatisfied.smv
-- specification  G (InitialNode1 ->  F VerifyCreditCard)  is true
-- specification  G (VerifyCreditCard -> ( F ReplyCreditCardNotOK xor  F CreateOrderBusinessObject))  is false
-- as demonstrated by the following execution sequence
Trace Description: LTL Counterexample 
Trace Type: Counterexample 
-> State: 1.1 <-
  InitialNode1 = TRUE
  ReceiveNewOrder = FALSE
  VerifyCreditCard = FALSE
  DecisionNode1 = undetermined
  ReplyCreditCardNotOK = FALSE
  DecisionNode2 = undetermined
  ActivityFinalNode1 = FALSE
  ConfirmOrderCancelation = FALSE
  ...
-> State: 1.2 <-
  InitialNode1 = FALSE
  ReceiveNewOrder = TRUE
-> State: 1.3 <-
  ReceiveNewOrder = FALSE
  VerifyCreditCard = TRUE
-> State: 1.4 <-
  VerifyCreditCard = FALSE
  DecisionNode1 = guard_DecisionNode1_DecisionNode2
-> State: 1.5 <-
  DecisionNode1 = undetermined
  DecisionNode2 = guard_DecisionNode2_ConfirmOrderCancelation
-> State: 1.6 <-
  DecisionNode2 = undetermined
  ConfirmOrderCancelation = TRUE
-> State: 1.7 <-
  ConfirmOrderCancelation = FALSE
  ActivityFinalNode2 = TRUE
-- Loop starts here
-> State: 1.8 <-
  ActivityFinalNode2 = FALSE
-> State: 1.9 <-
-- specification  G (ReplyCreditCardNotOK ->  F ActivityFinalNode1)  is true
...
\end{lstlisting} 

We can see that a corresponding counterexample is also produced by the NuSMV model checker with respect to the unsatisfied properties (see Listing~\ref{CounterExample}). By analyzing the counterexample, we can track down the inconsistency between the two models to know where the containment property has not been satisfied. As our approach presented in this paper mainly focuses on defining containment checking and automatic generating of the formal constraints and specifications as inputs for containment checking, we did not go deeper into the verification process or the automated analysis of the verification result yet (which we plan to do in our future work). 

\begin{figure}[htp]  
\centering 
\includegraphics[scale=0.80]{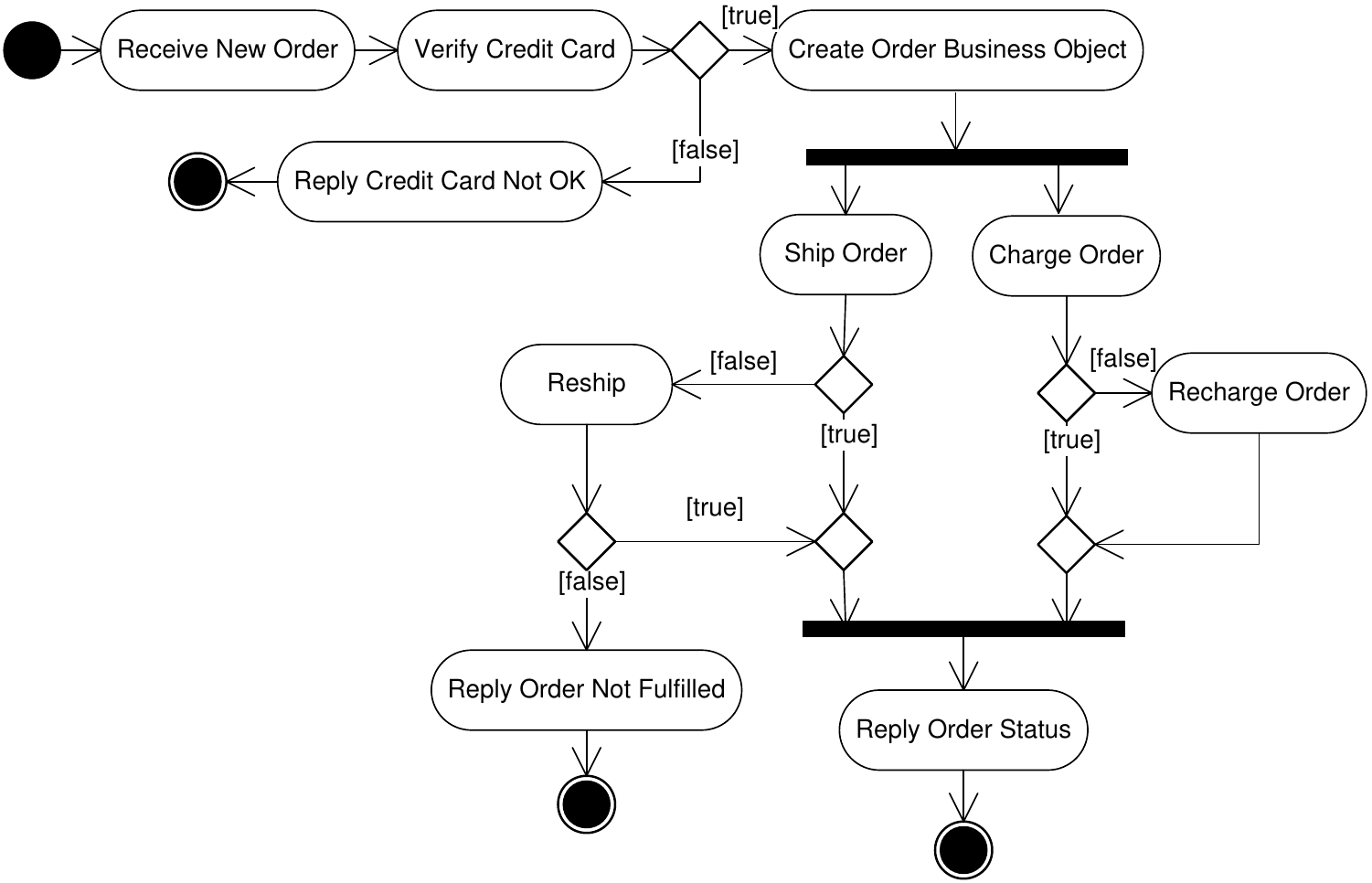}  	
\caption{A different version of the low-level UML activity diagram of the order processing system\label{fig:LowLevelOrderProcessingSatisfied}}  
\end{figure}

In Figure~\ref{fig:LowLevelOrderProcessingSatisfied} we illustrate another version of the low-level UML activity diagram of the order processing system that satisfies the containment checking constraints. The checking of the NuSMV description generated out of this model against the LTL formulas from Listing~\ref{GeneratedLTL} produces no unsatisfied properties, i.e., the NuSMV model checker returns true for all LTL formulas. This result indicates that the low-level behavior model is consistent with its high-level counterparts. 

\section{Discussion} \label{sec:discussion}

This section discusses some issues regarding the automated mappings of behavior models into formal specifications and properties used for containment checking. We have presented the automated mapping of basic constructs of UML activity diagrams like actions, activities, parallel nodes (fork and join), sequences, and branching nodes (decision and merge). At the current level of development, our approach has some limitations, for example, the automated mapping of other constructs of UML activity diagrams such as accept event actions, activity parameter nodes, and exception handling have not been considered yet. But we expect that the automated mapping of these constructs can be achieved in a similar manner with some extra effort. Furthermore, our current work has not incorporated the mapping of loops, data objects and object flows into LTL formulas. The UML 2 specification allows to complement the control flows of a UML activity diagram with data and object flows but does not allow a mix of object flows and control flows. Therefore, it is possible to specify data and object flows separately using the same techniques presented in this paper. Loops might introduce bigger challenges because a loop can produce deterministically or nondeterministically cyclic execution flows that is difficulty to transform to simple LTL constraints. We will investigate other variants of temporal logic such as bounded linear temporal logic~\cite{LatvalaB+2004} to overcome this challenge. These are, however, beyond the scope of this paper and part of our future endeavor. As future work, we also plan to apply our approach on realistic industrial systems. This will also be the basis to further investigate whether our proposed approach is able to support containment checking of software systems in practice. 

\section{Conclusion} \label{sec:conclusion}
In this paper, we present a novel approach for supporting containment checking of UML activity diagrams. On the one hand, the high-level UML diagrams, often used by business analysts and/or domain experts, are translated to LTL. On the other hand, the low-level counterparts, often resulting from various steps of refinement and enriching of these high-level models, are mapped onto formal NuSMV descriptions. By considering high- and low-level UML activity diagrams as inputs for containment checking and automatically transforming them into formal properties and descriptions, our approach can help lessening the gap and efforts for creating consistency constraints as many other approaches. The satisfaction of the NuSMV descriptions with respect to the LTL formulas, which can be verified using existing model checking tools, can denote the containment relationship between the high-level and low-level UML activity diagrams. 

However, it is difficult to covered whole UML specification in the scope of this paper. Therefore, we have mainly investigated fundamental elements and control structures of UML activity diagrams that are widely used for modeling software system behavior. Nevertheless, the same methods and techniques can also be adapted and applied for other control structures. Therefore, one of our future endeavors is to exploit our approach for various aspects of software behavior modeling such as data objects, object flows, error and exception handling, etc. Another aspect is to consider whether the performance our approach is reasonable in the context of typical software development environments.
\\[1ex] 
\noindent {\bf Acknowledgment.}
The research leading to these results has received funding from the Wiener Wissenschafts-, Forschungs- und Technologiefonds (WWTF), Grant No.\ ICT12-001.

\bibliographystyle{eptcs}
{
\bibliography{paper}
}
\end{document}